\documentclass{pasj01}
\usepackage{lscape,color,ulem}

\Received{$\langle$reception date$\rangle$}
\Accepted{$\langle$acception date$\rangle$}
\Published{$\langle$publication date$\rangle$}

\begin{document}

\title{The First MAXI/SSC Catalog of X-ray Sources in 0.7--7.0~keV}
\author{Hiroshi \textsc{TOMIDA}\altaffilmark{1}, 
Daiki~\textsc{UCHIDA}\altaffilmark{2}, 
Hiroshi~\textsc{TSUNEMI}\altaffilmark{2}, 
Ritsuko~\textsc{IMATANI}\altaffilmark{2}, 
Masashi~\textsc{KIMURA}\altaffilmark{3}, 
Satoshi~\textsc{NAKAHIRA}\altaffilmark{3}, 
Takanori~\textsc{HANAYAMA}\altaffilmark{4}, 
Koshiro~\textsc{YOSHIDOME}\altaffilmark{4}
}%
\altaffiltext{1}{Institute of Space and Astronautical Science, Japan Aerospace Exploration Agency, 3-1-1 Yoshinodai, Chuo, Sagamihara, kanagawa 252-5210 Japan}
\email{tomida.hiroshi@jaxa.jp}
\altaffiltext{2}{Department of Earth and Space Science / Project Research Center for Fundamental Science, Graduate School of Science, Osaka University, 1-1 Machikaneyama, Toyonaka, Osaka 560-0043 Japan}
\altaffiltext{3}{Human Spaceflight Technology Directorate, Japan Aerospace Exploration Agency, 2-1-1 Sengen, Tsukuba, Ibaraki 305-805 Japan}
\altaffiltext{4}{Department of Applied Physics, University of Miyazaki, 1-1 Gakuen Kibanadai-nishi, Miyazaki, Miyazaki 889-2192}
\KeyWords{Catalogs --- Surveys --- X-rays: general }

\maketitle

\begin{abstract}
We present the first source catalog of the Solid-state Slit Camera (SSC) of the Monitor of All-sky X-ray Image (MAXI) mission on the International Space Station, using the 45-month data from 2010 August to 2014 April in 0.7--7.0~keV bands.
Sources are searched for in two energy bands, 0.7--1.85~keV (soft) and 1.85--7.0~keV (hard),
the limiting sensitivity of 3 and 4~mCrab are \textcolor{black}{achieved}
and 140 and 138 sources are detected in the soft and hard energy bands, respectively.
Combining the two energy bands, 170 sources are listed in the MAXI/SSC catalog.
\textcolor{black}{All but 2 sources are identified with}
22 galaxies including AGNs,
29 cluster of galaxies,
21 supernova remnants,
75 X-ray binaries,
8 stars,
5 isolated pulsars, and
9 non-categorized objects.
Comparing the soft-band  fluxes  at the brightest end in our catalog with the ROSAT survey, which was performed about 20 years ago,
\textcolor{black}{10\% of  the cataloged sources are found to have changed the flux  since the ROSAT era.}

\end{abstract}

\section{Introduction}

The Monitor of All-sky X-ray Image (MAXI : \cite{2009PASJ...61..999M}) is the first astronomical payload aboard the \it International Space Station \rm (ISS).  MAXI has been  performing the monitoring observations since its first light in 2009 August.  The monitoring data of MAXI are open to the public at the RIKEN web site (http://maxi.riken.jp), where  the light curves of more than 380 X-ray sources are available.   \textcolor{black}{MAXI is equipped with} two types of cameras:  the Gas Slit Camera (GSC: \cite{2011PASJ...63S.623M}, \cite{2011PASJ...63S.635S}) and the Solid-state Slit Camera (SSC: \cite{2010PASJ...62.1371T}, \cite{2011PASJ...63..397T}).  The GSC is an array of position-sensitive gas proportional counters  with  beryllium windows with a total detection area of 5350\,cm$^2$. The SSC is an array of X-ray CCDs with a total detection area of 200\,cm$^2$.  Both have a fan-beam field of view (FOV)  and scan \textcolor{black}{almost the entire sky every} 92\,minutes. The FOVs of the GSC  and SSC are \timeform{1.D5}$\times$\timeform{160D} and  \timeform{1.D5}$\times$\timeform{90D}, respectively.

\textcolor{black}{The first all-sky catalog for X-ray sources was obtained by UHURU observations (\cite{1978ApJS...38..357F}) in the period between 1970 and 1973 and includes 339 sources in 2--6~keV.}
The Ariel-V  satellite, which was active between 1974 and 1980,  provided two catalogs: one  contains 142\,sources in 2--18~keV ($|b|>$\timeform{10D}, \cite{1981MNRAS.197..893M}) and the other,  109\,sources in 2--10~keV ($|b|<$\timeform{10D}, \cite{1981MNRAS.197..865W}).  The next comprehensive ones \textcolor{black}{were  published based on} the HEAO1 data between 1977 and 1979:   A1 catalog in 1--20~keV (842\,sources, \cite{1984ApJS...56..507W}), A2 catalog in 0.2--2.8~keV (114\,sources, \cite{1983ApJS...51....1N}), and A3 catalog in 13--180~keV (40\,sources, \cite{1984ApJS...54..581L}).   \textcolor{black}{All of these catalogs  were based on the data taken with non-focusing optics.}  ROSAT,  equipped with X-ray mirrors, performed all-sky survey during the first half year of the mission in 1990--1991.  Its energy range is in 0.1--2.4~keV.  The bright source catalog of ROSAT from the survey includes 18811\,sources (\cite{1999A&A...349..389V}), and the faint source catalog  does 105924\,sources (\cite{2000yCat.9029....0V}) in 0.1--2.4~keV.

There are two types of instruments to perform all-sky survey.  One has a relatively narrow FOV; UHURU (\timeform{1D}$\times$\timeform{10D}), Ariel-V (\timeform{0.D75}$\times$\timeform{12D}), HEAO1  (\timeform{2D}$\times$\timeform{8D} etc.) and ROSAT (\timeform{2D} circle) are of this type.  \textcolor{black}{They can take a long time to cover the entire sky, up to half a year in the case of ROSAT}, while individual sources are observed  for a relatively short period.  
\textcolor{black}{As a result, they can miss sources that happen to be in quiet phase during its covering period.}  \textcolor{black}{The other has a much wider FOV, like MAXI.
They periodically scan a large area of, or even the entire, sky with a relatively short interval, for example, 92 minutes with MAXI.   The source intensities  listed in our MAXI catalog are accordingly the average ones over 3.7 years of our survey period presented in this paper. Hence, MAXI is far less likely to miss sources that fall in quiescence for a relatively short period than the other type of all-sky survey instruments.}
   
All-sky survey in the X-ray band is a powerful tool to investigate the  entire spectrum of  high-energy phenomena in the universe since most of  X-ray sources show strong time variabilities in various time scales, from  a milli-second to years. \textcolor{black}{MAXI has a potential to upgrade the existing all-sky X-ray source catalog.}   \citet{2011PASJ...63S.677H}  presented the first GSC catalog in the 4-10~keV band that contains 143\,sources at high galactic latitudes ($|b|>$\timeform{10D}), \textcolor{black}{using the first 7 months of the monitoring data}.  \citet{2013ApJS..207...36H} later updated the GSC catalog,  using the 37-month data that contains more than 500 sources.  The limiting sensitivity reaches 7.5$\times$10$^{-12}$ erg cm$^{-2}$ s$^{-1}$ (0.6~mCrab) in the 4--10 keV band. 

Since the MAXI/GSC employs a similar detector to those of UHURU and HEAO1/A1,  the GSC catalog covers the energy range  between a few keV  and a few tens of keV,  \textcolor{black}{which is above that} of ROSAT.  Here, we report the MAXI/SSC catalog that covers the energy range of 0.7--7.0~keV.  We describe the data reduction of the MAXI/SSC data in section 2 and  \textcolor{black}{method of the source detection with} the SSC data in section 3.  In section 4, we present the SSC source list and discussions.  Summary is presented in section 5.

\section{Data reduction}

The SSC consists of 2\,units: SSC-H and SSC-Z.  
\textcolor{black}{The two units are identical and are composed of 16 CCDs
with a total detection area} of 100\,cm$^2$ per unit.  SSC-H  views the forward direction of the ISS,  while SSC-Z  does the zenith (anti-earth) direction.  Each is equipped with a fan-beam collimator of \timeform{1.D5}$\times$\timeform{90D}.  CCDs  work in  the pixel-sum mode that functions as one-dimensional imager along the long direction of the collimator.  The details are given in the literatures (\cite{2010PASJ...62.1371T}, \cite{2011PASJ...63..397T})

\begin{figure}
 \begin{center}
   \includegraphics[width=8cm]{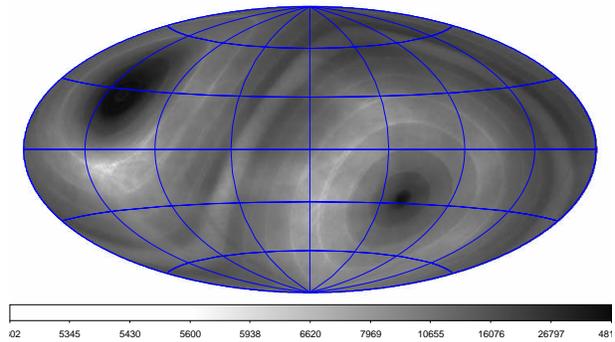}
 \end{center}
 \caption{Exposure map (effective area $\times$ exposure time) of the SSC data that we used for the catalog production in the galactic coordinates.
   The  gray-scale is in logarithmic scale.
   The striped pattern was caused by gaps between the CCDs in the X-ray cameras. }
 \label{exposure_map}
\end{figure}

\textcolor{black}{MAXI has been in operation since 2009 till present.
However, we use the data taken between 2010 August 1 and 2014 April 30 only for the catalog presented in this paper, discarding the data before and after for the following reasons.}
The SSC event data before 2010 August were heavily contaminated with dark-level saturated pixels due to sun-light leak.  It was thus difficult to estimate the correct effective area. \textcolor{black}{The data taken after 2014  April were seriously affected by flickering pixels }\textcolor{black}{because gradual increase of the dark current since the launch had reached an unacceptable level by then, which generated many flickering pixels.}  

We exclude the data taken around the geomagnetic pole at high latitude and the South Atlantic Anomaly region,  which suffer from many charged-particle events.  We also exclude the data obtained 
during day-time.
 In order to exclude the charged-particle events, we accept grade-0 (single pixel) events as X-rays below 1.85\,keV, and grade-0, 1, and 2 (single pixel and two-pixel split) events above 1.85\,keV just as done by \citet{2013PASJ...65...14K}.  

\textcolor{black}{The SSC is equipped with the radiator and the peltier cooler, which cools the CCD around \timeform{-60D}C. Since the temperature of one CCD (SSC-Z/CCDID=0) is rather high due to insufficient function of the peltier cooler attached on it, we exclude it from our analysis.  }  \textcolor{black}{Due to the thermal noise,  the events with the energy  lower than 0.7\,keV are practically unusable and are discarded.}  Since the collimator is made of  copper that generates Cu-K lines on all the CCDs, we set the  upper energy limit of the SSC data to be 7.0\,keV.   Consequently, we  accept the X-ray events between 0.7\,keV and 7.0\,keV.

Figure~\ref{exposure_map} is the exposure map obtained through the above-mentioned screening method. \textcolor{black}{The daily observation coverage of the MAXI/SSC was limited to smaller than 40\% of the entire sky, considering the FOV ($\pm$45$^{\circ}$) and the day-time data discarded.} The ISS orbital plane shows a precession in every 70\,days (the inclination angle of the ISS orbit is \timeform{51.D7}).    Accordingly, the MAXI/SSC  covers the entire sky in a yearly observation.   \textcolor{black}{The region around the celestial north pole is more frequently observed than other regions and has a deep exposure (figure~\ref{exposure_map})}.  The ISS moves around the Earth  at the almost  constant attitude.  This means that MAXI  always  looks at sky and never sees the dark-Earth nor bright-Earth,  which is in stark contrast to other satellites in a low Earth orbit like Suzaku (\cite{2008PASJ...60S..11T}).  

\section{Analysis}
\textcolor{black}{After obtaining the screened X-ray data,  we take two steps to detect X-ray sources.} The first one is the image production for the entire sky.  Then the second one is  source finding.  We execute these  steps for two energy bands, 0.7--1.85\,keV (soft band) and 1.85--7.0\,keV (hard band), where the energy of 1.85\,keV corresponds to the boundary of the grade selection (see the previous section).  Since the low-energy limit of the MAXI/GSC detection is 2\,keV, the soft-band data of the SSC provide us with a view of the sky that is invisible with the MAXI/GSC.  The hard-band data are similar to that of the MAXI/GSC  except that the working time is not always the same due to the observation conditions.  We can compare sources detected in our hard-band data with those of the GSC catalogs (\cite{2011PASJ...63S.677H}, \cite{2013ApJS..207...36H}) to check the consistency, \textcolor{black}{as performed and described in section 3.4.}
 

\subsection{Image production}

\begin{figure}
 \begin{center}
   \includegraphics[width=8cm]{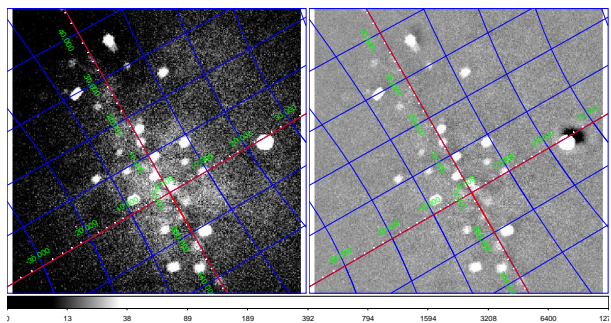}
 \end{center}
 \caption{Sample images used for source finding (Galactic center region).
   The left panel is the raw image, and the right one is the background-subtracted one.
   The intensity of both images are in  the logarithmic scale,
   but the scale units  are different.
 }
 \label{image_sex}
\end{figure}

We create 48 square images,  which cover the entire sky \textcolor{black}{as a whole}.  The center position of each square is selected by using Healpix (Hierarchical Equal Area isoLatitude Pixelization: \cite{2005ApJ...622..759G}).  We select  each image to be \timeform{67D.5} square  to ensure that it covers one Healpix segment.  The left panel of figure~\ref{image_sex} is a sample of  the square images, near the galactic center.  It is neither background-subtracted nor exposure-corrected.  We divide each image into 300$\times$300 pixels.  The pixel is \timeform{0D.225} square that is small enough not to affect the shape of the point spread function (PSF) of the SSC (\timeform{1D.5} in FWHM).  Each square image overlaps its adjacent images.  Therefore, when we find a source in  an image, \textcolor{black}{we check that it is not doubly counted  because of this overlap.} When a source is very close to the segment boundary, we  make visual inspection.

\begin{figure}
 \begin{center}
   \includegraphics[width=8cm]{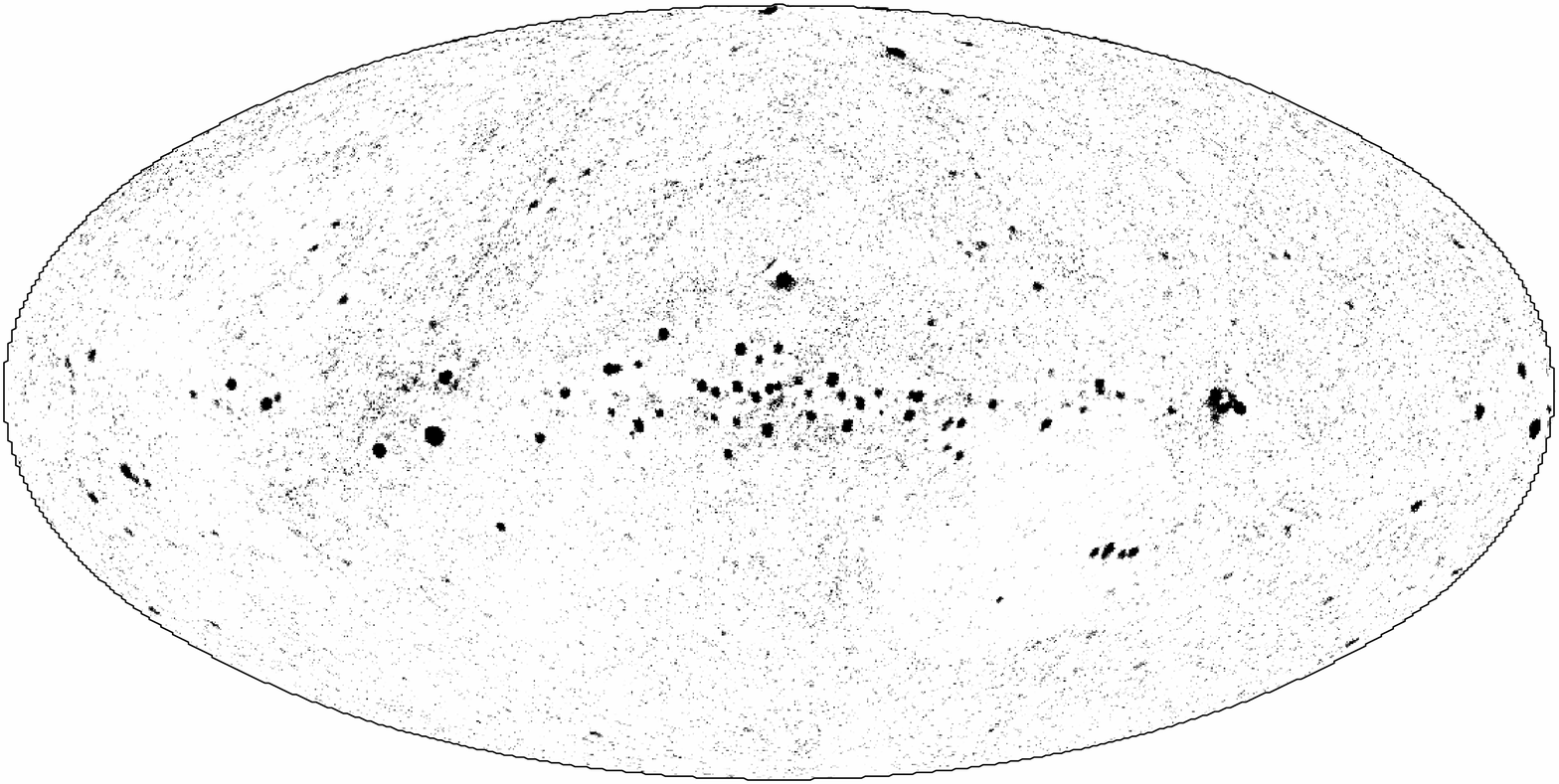}
   \includegraphics[width=8cm]{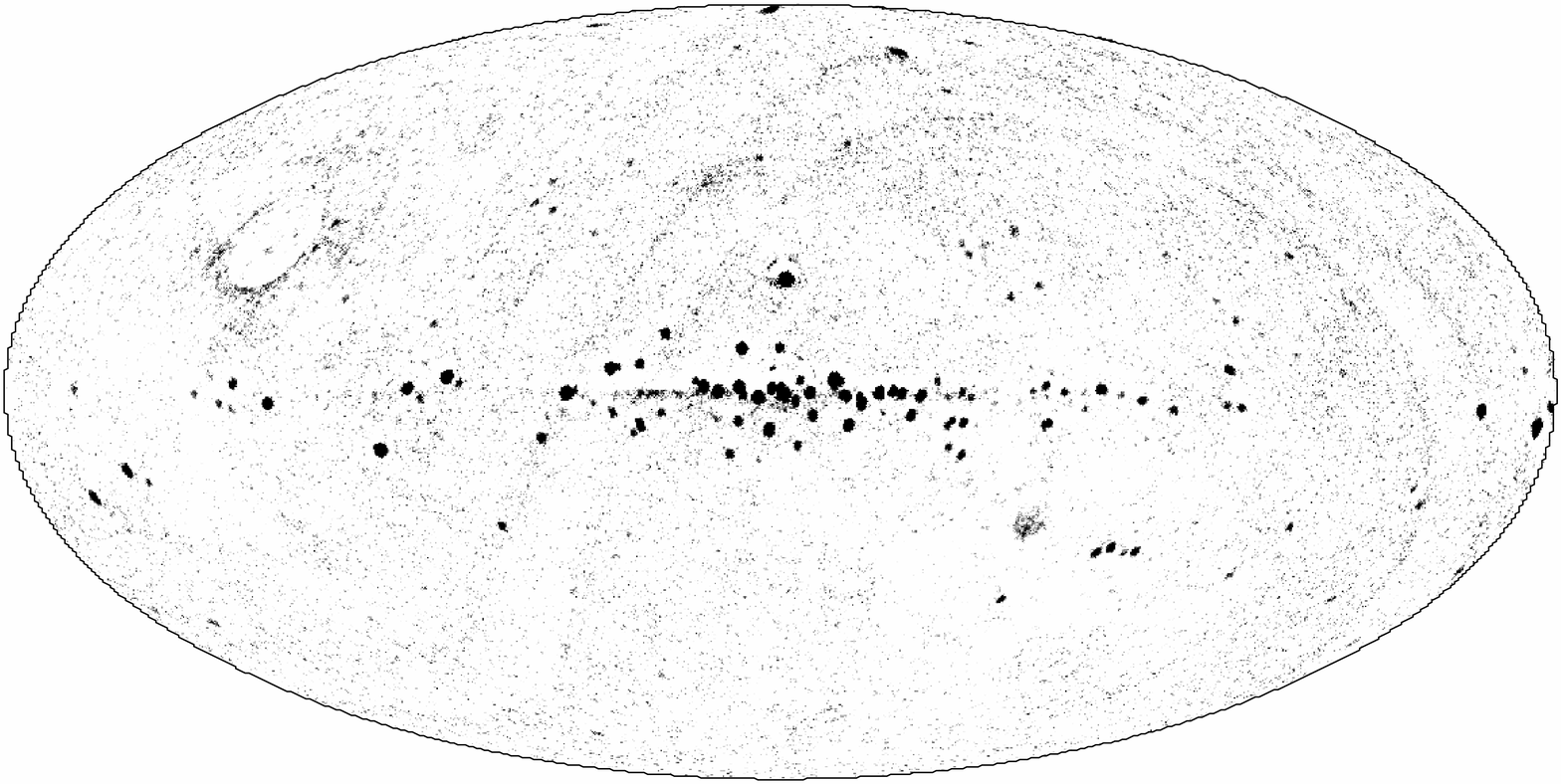}
 \end{center}
 \caption{All-sky map of SSC data in the galactic coordinates in the soft (left) and hard (right) energy
    bands.
   The extended component determined with SExtractor is  subtracted.}
 \label{all-sky_map}
\end{figure}

\subsection{Source finding}

\textcolor{black}{We  employ the SExtractor V2.5.0 (Source Extractor : \cite{1996A&AS..117..393B}) for source detection in the images produced as described in the previous section.}  
\textcolor{black}{In SSC images, most of detected sources are expected to be point-like.  The SExtractor requires the following three parameters to be carefully set to detect point-like sources accurately.}  The parameter \it BACK\_SIZE \rm is used to subtract the background (extended) component \textcolor{black}{and it is tuned to the size of the PSF.} \it DETECT\_THRESHOLD \rm is used to determine the significance of the source candidate.  \it DETECT\_MINAREA \rm is the minimum number of adjacent pixels above \it DETECT\_THRESHOLD, \rm which is  set to a slightly smaller value than the pixel number in the PSF of the MAXI/SSC. \textcolor{black}{Table~\ref{sextract} summarizes these three parameters.}

The right panel of figure~\ref{image_sex} is a background-subtracted image created  with SExtractor from the left panel. \textcolor{black}{With the procedure described above}, we create 48 background-subtracted images and combine them to obtain the  entire-sky image shown in figure~\ref{all-sky_map} in the galactic coordinates.    We  have extracted 814 and 754 sources,  using SExtractor, in the soft- and hard-band images, respectively.  
\textcolor{black}{Then we check the Healpix segment, to which each source belongs, to avoid double counts.} Some sources  are located very close to  a segment boundary.   When two sources are located within \timeform{0D.1}, we select one of them.  In this way, we filter out duplicated sources and obtain 350 and 347 sources as candidates for the soft and hard bands, respectively.

\begin{table}
  \tbl{The parameters for SExtractor in the SSC source-finding}{%
  \begin{tabular}{lc} \hline
    \it BACK\_SIZE \rm         & 10  \\
    \it DETECT\_THRESHOLD \rm  & 1.0 \\
    \it DETECT\_MINAREA \rm    & 18 \\ \hline
  \end{tabular}}
  \label{sextract}
\end{table}


The next step of the catalog creation is  \textcolor{black}{validation} of the source candidates.  Our method for the  validation is rather simple.  We compare the number of events in a candidate region with that of the background region.   Each candidate region is a circle with a radius of \timeform{1D.5}, and the background region is an contiguous annular region with the  outer radius of  \timeform{3D.0}.  We then define the significance level, $S_D$, as 
\begin{equation}
 S_{D} = (N_s - N_b \times S_{c})/ ( \sqrt{N_s + N_b \times S_{c}^{2}} ),
\end{equation}
where $N_s$ is the number of events in  a candidate region, $N_b$ is that of the background region, and $S_c$ is the ratio of the areas of the candidate  to  the background regions.
\textcolor{black}{We define $S_D > $5.0  as the threshold for significant detection.}

 \textcolor{black}{Crowded regions} \textcolor{black}{ must be treated with caution.}  When a region overlaps  those of other sources, we simply exclude the overlapping region.  However, when the nearby source is determined to be  insignificant (or fake), we recalculate the source and/or background region.  We iterate this process to calculate the  significance $S_D$ until no fake source is obtained.  In this way, we  extract the list of the true sources from candidates.

\textcolor{black}{For each of the true sources,} we calculate the center of gravity of  events in the source region and  assign it as the candidate source position.  Then, we re-define the circle of the source region and calculate the center of gravity again.  We iterate this procedure until the shift of the source position becomes  smaller than \timeform{0D.02}.  Since we do not take into account the potential effects from nearby sources, the position accuracies may be worse than \timeform{0D.02}, particularly in  crowded regions.

The source flux is evaluated from the number of events in the background-subtracted image.  We then compare it with the number of events in the Crab simulation, where we assume   the photon index of 2.1,  the interstellar absorption N$_H$ of 3.4$\times$10$^{21}$ cm$^{-2}$ with the solar abundance\footnote{The spectral parameters are adopted from the INTEGRAL General Reference Catalog (ver. 40).  http://www.isdc.unige.ch/integral/catalog/40/catalog.html}, and
the photon flux density  at 1.0~keV  of 10\,photon sec$^{-1}$ cm$^{-2}$ keV$^{-1}$.  We perform the simulation,  using XSPEC version 12.8.2 and the spectrum response files created in the same way as those distributed at the MAXI web site. \textcolor{black}{The flux is expressed in units of erg sec$^{-1}$ cm$^{-2}$.}

In our SSC data \textcolor{black}{of the Crab observations,} we find the  flux to be 8.21$\times$10$^{-9}$ and 1.824$\times$10$^{-8}$ erg sec$^{-1}$ cm$^{-2}$ for the soft and the hard bands, respectively.  On the other hand, our simulation of the Crab gives 9.34$\times$10$^{-9}$ and 1.807$\times$10$^{-8}$ erg  sec$^{-1}$ cm$^{-2}$, respectively.  \textcolor{black}{Hence, the  observed results with the SSC are 0.88 and 1.01 times that of  the  simulation.}  \textcolor{black}{Although this difference depends partially on the systematic error of the SSC normalization, the differences mainly depends on the spectral model. The Crab model that we employed was determined with the INTEGRAL mission, of which the energy range is $>$3~keV. Then, the  discrepancy in the soft band is larger than that of the hard band.}

\subsection{False detection}

\textcolor{black}{In order to study how probable the detected source candidates originate in the potential background fluctuations, as opposed to true sources,}
we search for negative peaks in the 48 images  the same as  by \citet{2011PASJ...63S.677H}.  The procedure of the negative-peak search is the same  as in the positive-peak search.   Consequently, we detected 21 candidates with $S_D$ $>$ 5.0 as negative peaks.  We find all of them to be close to bright sources. Indeed, in the background-subtracted images,  some dark regions close to the bright source are visible  (see the right panel of figure~\ref{image_sex}, where Sco X-1 is in the right edge).  \textcolor{black}{These dark regions are most likely to be} mainly caused by inappropriate background estimation.
The PSF of the SSC is a box-car shape of \timeform{2D.0}$\times$\timeform{3D.0} in the single scan image (figure 10 in \cite{2010PASJ...62.1371T}).  The SSC scans each source many times, while the scan direction  is not always randomly distributed but is  determined preferentially by the ISS orbit. \textcolor{black}{As a result, the PSF of the SSC is clearly skewed from a cirlce even in the integrated image for 4 years of the data, although the PSF is expected to be considerably smoothed out after integration.} We set the parameter \it BACK\_SIZE \rm to 10 (section 3.2 and table 1).  However, this is not  appropriate for some directions of scan, particularly in the  vicinity of bright sources.  \textcolor{black}{We find that all of the 21 negative  peaks are  visible} only in the background-subtracted images. \textcolor{black}{Therefore, they all must be false detection.}  We also find that $S_D$ $>$ 5.0 is large enough to avoid the false detection due to the background fluctuation. 
\textcolor{black}{Applying this threshold of $S_D > 5.0$, more than half of the sources in the source lists are filtered out, and 154 and 154 sources remain as significant in the soft and hard bands, respectively.}

 \textcolor{black}{For all the significant sources,} we perform  visual inspection and find some false detections, \textcolor{black}{as described below}.  In the hard-band image of figure~\ref{all-sky_map}, we see some bright regions,  \textcolor{black}{which are suspected} artificial structures: stripes parallel to the celestial equator and a region around the celestial south pole.  From our visual inspection, we find \textcolor{black}{23} sources that must originate from the fluctuation of the background.  Then,  \textcolor{black}{the number of the sources are reduced to} \textcolor{black}{140} and 138  for the soft and hard bands, respectively, \textcolor{black}{after the 23 sources are excluded.} 

 By comparing the source positions in the two energy bands, we find that \textcolor{black}{108} sources are  \textcolor{black}{common in both the energy bands,  where their positional separations are smaller than \timeform{0D.5}.   Consequently, the final SSC catalog contains 170 unique sources.}   Table 2 summarizes the number of source candidates at each source-finding step. 

\begin{table}
  \tbl{The number of source candidates at each step in creating the SSC catalog.}{%
  \begin{tabular}{lcc} \hline
    \multicolumn{1}{l}{} &  \multicolumn{1}{l}{soft} & \multicolumn{1}{c}{hard} \\ \hline
    first candidates  by Sextractor & 814 & 754 \\
    duplicate check       & 350 & 347 \\
    significance ($S_D$) $>$ 5.0  & \textcolor{black}{154} & \textcolor{black}{154} \\
    visual inspection     & \textcolor{black}{140} & 138 \\ \hline
  \end{tabular}}
  \label{source_number}
\end{table}

\subsection{Source identification}

 X-ray sources usually show strong variabilities in various time scales. Hence, we compare the SSC source catalog with that obtained by the MAXI/GSC rather than those obtained in the past by other satellites.  We employ the source list in the MAXI web-site and the MAXI/GSC high-galactic source catalog (GSC Catalog: \cite{2013ApJS..207...36H}).  

First,  we focus on the hard-band sources at  high galactic latitudes,  \textcolor{black}{because the energy bands and  periods of observations are almost identical  between the SSC and GSC catalogs  except for the detection limits.}  We find that all of the 58 hard sources at $|b|>$10.0 in the SSC catalog have a counter part in the GSC catalog.  This fact  confirms that our analysis procedure is appropriate.

 Then,  for the SSC soft band, we employ the ROSAT Bright Source Catalog (ROSAT/BSC: \cite{1999A&A...349..389V}) for source identification, \textcolor{black}{as the GSC is insensitive to the energy band.}  The ROSAT/BSC does not include  highly extended objects like supernova remnants (SNR) and cluster of galaxies.  Therefore, we also employ the SNR catalog (\cite{2014BASI...42...47G}) and REFLEX galaxy cluster survey catalog (\cite{2004A&A...425..367B}).  Taking into account the positional accuracy of the SSC, we set the \textcolor{black}{upper threshold for the angular distance for identification  to be  \timeform{0D.5}. }

\section{Results and Discussion}

\begin{figure}
 \begin{center}
   \includegraphics[width=8cm]{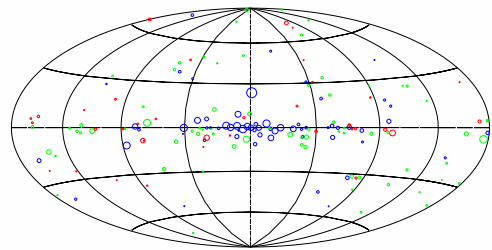}
 \end{center}
 \caption{
   Distributions of the MAXI/SSC sources in the galactic coordinates.
   The radii of circles are proportional to log(flux),
   and colors  represent the softness ratio ($SR$: soft-band flux divided by hard-band flux in table \ref{ssc_catalog});
   red , green, and blue marks are for the sources with  $SR$$<$0.5,  $0.5 \leq SR \leq 1.0$,
   and $SR$$>$1.0, respectively. }
 \label{image_sorce_dist}
\end{figure}

We summarize the MAXI/SSC catalog in table~\ref{ssc_catalog},  in which sources with S$_D >$5.0 are listed.
The columns  in the table are 
(1) identification number,
(2) MAXI-SSC name,
(3) Right Ascension (degree in J2000),
(4) Declination (degree in J2000),
(5)(6) detection significances in the soft  and  hard bands, respectively,
(7)(8) flux in the soft  and  hard bands, respectively,
(9) source identification,
and
(10)  counterpart in the MAXI/GSC catalog.
The source locations in the galactic coordinates are plotted in figure~\ref{image_sorce_dist}.

\subsection{Source category}

In table~\ref{source_category}, we summarize the \textcolor{black}{statistics of the source  identifications,  referring} to the 9\,th column of table~\ref{ssc_catalog}.   \textcolor{black}{The MAXI/SSC catalog contains  more binary sources than the MAXI/GSC catalog,  because} the MAXI/SSC catalog includes sources at low galactic latitudes.  The high sensitivity of the SSC in the soft band  yields detection of many SNRs.

\begin{table}
  \tbl{Source category of the SSC catalog}{%
  \begin{tabular}{lc} \hline
    \multicolumn{1}{l}{Category} & \multicolumn{1}{c}{Number of sources} \\ \hline
    Galaxies/AGNs & 22\\
    Cluster of Galaxies & 29\\
    SNRs & 21\\
    X-ray binaries & 75 \\
    stars & 8 \\
    Isolated pulsar & 5 \\
    unknown/no identification & 11 \\ \hline
  \end{tabular}}
  \label{source_category}
  \begin{tabnote}
    The number of sources in table~\ref{ssc_catalog} is 170,
    while the sum of the number\textcolor{magenta}{s} in this list is 171.
    This is because the source ID=43 in table~\ref{ssc_catalog} consists of two sources: Vela pulsar and SNR.
  \end{tabnote}
\end{table}

\subsection{Position accuracy}

\begin{figure}
 \begin{center}
   \includegraphics[width=60mm,angle=270]{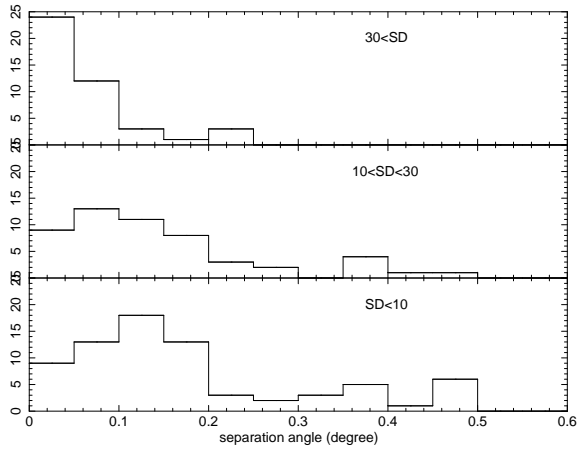}
 \end{center}
 \caption{Source distribution as a function of the angular separation between the SSC position and that of the counter part.
   The top panel is the distribution of $S_D > $30, middle is of 30$\ge  S_D >$10, and the bottom is of 10$ \geq S_D$.
   For sources detected in both  the soft and hard bands, the smaller separation angle is  adopted.
 }
 \label{position}
\end{figure}

From the source identification result, we estimate the positional errors of the MAXI/SSC as a function of $S_D$ (\textcolor{black}{the columns 5 and 6} in table \ref{ssc_catalog}). \textcolor{black}{Figure \ref{position} shows the distribution of the angular separation between the MAXI/SSC sources and the identified sources for three intervals of detection significance.} As can be seen, sources with higher significance have smaller angular separations. \textcolor{black}{In the top panel, relevant to sources with $S_D > 30$}, the statistical errors are smaller than the systematic errors. \textcolor{black}{When the significance is high enough, $S_D > 30$ (top panel), the statistical errors must be dominant, compared with the systematic errors.  Thus, we estimate the systematic error of the source positions, using the 43 sources with $S_D > 30$.  Out of the 43 sources,} three sources  have the separation $>$\timeform{0D.2}:   GX~3+1, Pup~A, and LMC~X-1.  GX~3+1 is near the galactic center,  and LMC~X-1 is in the Large Magellanic Cloud.  These two are very close to  bright sources,  which affect the accuracies of source positions.  Pup~A is an extended SNR that is on the edge of the large SNR, Vela-SNR,  which also affects the positional accuracy.  Therefore, these three sources are  exceptional.   The separation angle of any of the other 40 sources than these three sources with $S_D > 30$  is smaller than \timeform{0D.2}.  We conclude that the systematic error of the position determination for the MAXI/SSC is  at largest \timeform{0D.2}.

\subsection{log$N$-log$S$ relation}

\begin{figure}
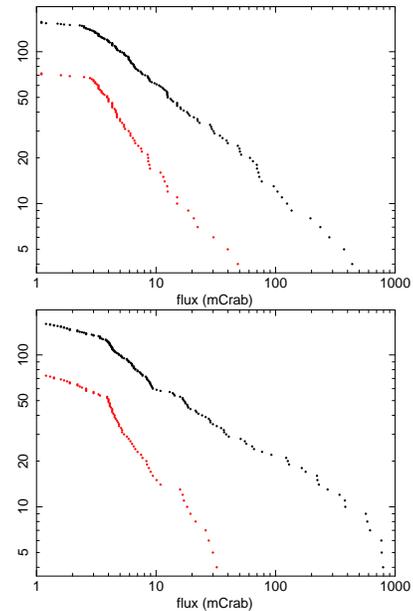

 \begin{center}
 \includegraphics[width=40mm,angle=270]{logN_logS_soft.ps}
 \includegraphics[width=40mm,angle=270]{logN_logS_hard.ps}
 \end{center}
 \caption{Plot of log$N$-log$S$ relation for the soft-band (left panel)  and  hard-band (right-panel) sources.  The black dots are sources in the entire sky,  and the red dots are  those  at high galactic latitudes ($|b| >$ 10 deg).  The source flux is given in units of mCrab.  
}
 \label{logNlogS}
\end{figure}

The log$N$-log$S$ relation from  a survey observation provides us with the fundamental information of the source population, where $N$ is the number of sources brighter than the flux $S$.
Figure \ref{logNlogS} shows the  one based on the SSC catalog.  Red marks in figure~\ref{logNlogS} are the sources at high galactic latitude ($|b|>$\timeform{10D}) \textcolor{black}{: this distribution is well fit by a straight line with a slope of 1.5 above} 3~mCrab and 4~mCrab for the soft and hard band plots, respectively.  The break points should correspond to the sensitivity limit of the SSC catalog.   Consequently, the limiting sensitivities of the MAXI/SSC catalog are 3~mCrab and 4~mCrab for the soft and hard bands, respectively.

\subsection{Comparison with the ROSAT era}

Since the MAXI/SSC catalog in the soft band covers the all sky below 2\,keV, we compare it with the ROSAT \textcolor{black}{all-sky} catalog.  We find that 18  out of 140 sources in the SSC soft-band catalog have no counterpart in the ROSAT/BSC (\cite{1999A&A...349..389V}).  We find that 9  out of the 18 \textcolor{black}{sources with no counterpart are  highly} extended sources (8~SNRs including Cygnus supper bubble and M87/Virgo).  They  were clearly detected by ROSAT, \textcolor{black}{although are not included in the BSC, which does not list highly extended sources.}   Six out of the remaining 9 sources are new transients after the ROSAT survey:
MAXI~J0556$-$332\ (\cite{2013PASJ...65...58S}), 
MAXI~J1836$-$194\ (\cite{2011ATel.3611....1N}), 
HETE~J1900.1$-$2455\ (\cite{2005ATel..516....1V}), 
Swift~J1753.5$-$0127\ (\cite{2005ATel..546....1P}), 
MAXI~J1910$-$057/Swift~J1910.2$-$0546\ (\cite{2012ATel.4140....1U}, \cite{2012ATel.4139....1K}), 
and GRS~1915+105\ (\cite{1994ApJS...92..469C}). 
The rests are 4U~1608$-$52 and two unidentified sources.  
4U~1608$-$52 is an Atoll source,  which shows a transition (\cite{1989A&A...225...79H}). \textcolor{black}{It was likely to be} in the quiescent phase at the time of the ROSAT survey.   The two unidentified sources (ID=137 and 141 in table~\ref{ssc_catalog}) are on the galactic plane.  \textcolor{black}{Hence, we can not exclude the possibility that they} may be fluctuations of the galactic ridge emission and are mistakenly identified as sources.

Next, we search the SSC catalog for the brightest 50 ROSAT sources,  and  find that  35 sources  have a counterpart.  
 \textcolor{black}{Table~\ref{SSCmissed} summarizes the nature of the other 15 sources with no counterpart in the SSC catalog,  taken from the SIMBAD\footnote{http://simbad.u-strasbg.fr/simbad} database}. 
 Eight out of the 15 sources are transient,  
and 6 out of the 8  have not been detected with the MAXI/GSC,  either.
The 6 sources are 
1RXS~J074833.8$-$674505,
1RXS~J112623.5$-$684040,
1RXS~J173413.0$-$260527, 
1RXS~J173602.0$-$272541,
1RXS~J175840.1$-$334828, and
1RXS~J215852.2$-$301338.
They  had been active in the ROSAT era,  but were quiescent during the MAXI survey period.  
 The other two out of the 8 were detected with the MAXI/GSC: 1RXS~J152040.8$-$571007 and 1RXS~J170248.5$-$484719.   \textcolor{black}{This difference in detection is} probably due to the difference in the coverage periods between the GSC and the SSC.   We should note that GX339$-$4 was in an almost quiescent phase during the period  of the SSC  survey, used in this paper.  It  became bright  in 2014  October\ (\cite{2014ATel.6649....1Y}).  
Another 4 of the 15 are white dwarfs  
(1RXS~J064509.3$-$164241, 
1RXS~J131621.4+290555,  
1RXS~J150209.2+661220, and 
1RXS~J200542.0+223955). 
White dwarfs are bright only below 0.7\,keV, which is below the energy band of  the SSC.   
One source out of the four white dwarfs is a cataclysmic variable of AM Her type, 1RXS~J200542.0+223955. 
\textcolor{black}{AM Her-type cataclysmic variables are known to emit hard X-rays and 1RXS~J200542.0+223955
is not an exception. However, the hard X-ray flux from it measured with the Ginga/Lac (\cite{1995MNRAS.273..742B}) is far below the detection limit of the SSC, 
which is consistent with our non-detection.}
Another undetected source, 1RXS~J052502.8$-$693840, is SNR N132D in Large Magellanic Cloud,  and is located close to a bright source LMC~X-1.  
MAXI/SSC can not separate N132D from LMC~X-1 due to the small angular distance.  
Another undetected source, 1RXS~J233733.3+462736, is an RS CVn type star, which shows  large X-ray flares in a time scale of hours.
ROSAT/BSC might have detected  a flare(s).
The fluxes  in this MAXI/SSC catalog are the average values for 4-years observations. Therefore, the SSC catalog  may have missed short (hour- or day-scale) transients  like 1RXS~J233733.3+462736. The short transient search with the MAXI/SSC will be presented in  a future paper.


\begin{table*}
  \tbl{Bright ROSAT sources that the SSC did not detect.}{%
  \begin{tabular}{llll} \hline
    \multicolumn{1}{l}{BSC name (1RXS~)} &  \multicolumn{1}{l}{identification} & \multicolumn{1}{l}{class* } & \multicolumn{1}{l}{comment} \\ \hline
J052502.8-693840 & N132D        & C &  SNR close to LMC~X-1  \\
J064509.3-164241 & H1504+65/Sirius-B     & B &  White Dwarf    \\
J074833.8-674505 & EXO~0748$-$676 & A &  X-ray burster     \\
J112623.5-684040 & GRS~1124$-$683 & A &  Black hole \\ 
J131621.4+290555 & HZ43         & B &  White Dwarf      \\
J150209.2+661220 & H1504+65     & B &  White Dwarf     \\
J152040.8-571007 & Cir~X-1      & A &  Neutron Star. Detected with MAXI/GSC   \\
J161741.2-510455 &              & E &     \\
J170248.5-484719 & GX339$-$4      & A &  Black hole. Detected with MAXI/GSC   \\
J173413.0-260527 & KS~1731$-$260  & A &  X-ray burster.  \\
J173602.0-272541 & GS~1732$-$273/KS~1732$-$273 & A &  Black hole \\ 
J175840.1-334828 & 4U~1755$-$338  & A & Black hole \\  
J200542.0+223955 & QQ Vu        & B &  CV of AM Her type  \\
J215852.2-301338 & QSO B2155$-$304/1H~2156$-$304  & A &  BL Lac    \\  
J233733.3+462736 & lam And      & D &  Variable of RS CVn   \\   \hline %
  \end{tabular}}
  \label{SSCmissed}
  \begin{tabnote}
* The third column indicates A:transient source, B:very soft source(white dwarf), C:source confusion, D:flare star, E:others. 
  \end{tabnote}
\end{table*}


\section{Summary}

We present the first MAXI/SSC catalog produced from an unbiased X-ray survey for 45 months in the 0.7--1.85~keV (soft) and 1.85--7.0~keV (hard) bands.  The limiting sensitivity of 3 and 4~mCrab are achieved and 140 and 138 sources are detected for the soft and hard bands, respectively.  Combining the data in the two bands,   the MAXI/SSC \textcolor{black}{catalog contains 170 unique sources.  The breakdown is} 22 galaxies including AGNs, 29 cluster of galaxies, 21 supernova remnants, 75 X-ray binaries, 8 stars, 5 isolated pulsars, 9 non-categorized objects, and 2  unidentified objects.  Comparing the soft-band catalog with the ROSAT survey, which was performed about 20 years ago, we find that  roughly 10\% of sources  have changed in intensity.

\begin{ack}
This research has made use of the MAXI data provided by RIKEN, JAXA and the MAXI team. 
Some of the results in this paper have been derived using the HEALPix (\cite{2005ApJ...622..759G}) and the SExtractor (\cite{1996A&AS..117..393B}).  
This research is partially  supported  by  the  Ministry  of  Education,  Culture, Sports, Science and Technology (MEXT), Grant-in-Aid No.23000004 and No.24103002.
\end{ack}
\bibliographystyle{apj}

\begin{thebibliography}{}
\expandafter\ifx\csname natexlab\endcsname\relax\def\natexlab#1{#1}\fi

\bibitem[{{Beardmore} {et~al.}(1995){Beardmore}, {Ramsay}, {Osborne}, {Mason},
  {Nousek}, \& {Baluta}}]{1995MNRAS.273..742B}
{Beardmore}, A.~P., {Ramsay}, G., {Osborne}, J.~P., {et~al.} 1995, \mnras, 273,
  742

\bibitem[{{Bertin} \& {Arnouts}(1996)}]{1996A&AS..117..393B}
{Bertin}, E., \& {Arnouts}, S. 1996, \aaps, 117, 393

\bibitem[{{B{\"o}hringer} {et~al.}(2004){B{\"o}hringer}, {Schuecker}, {Guzzo},
  {Collins}, {Voges}, {Cruddace}, {Ortiz-Gil}, {Chincarini}, {De Grandi},
  {Edge}, {MacGillivray}, {Neumann}, {Schindler}, \&
  {Shaver}}]{2004A&A...425..367B}
{B{\"o}hringer}, H., {Schuecker}, P., {Guzzo}, L., {et~al.} 2004, \aap, 425,
  367

\bibitem[{{Castro-Tirado} {et~al.}(1994){Castro-Tirado}, {Brandt}, {Lund},
  {Lapshov}, {Sunyaev}, {Shlyapnikov}, {Guziy}, \&
  {Pavlenko}}]{1994ApJS...92..469C}
{Castro-Tirado}, A.~J., {Brandt}, S., {Lund}, N., {et~al.} 1994, \apjs, 92, 469

\bibitem[{{Forman} {et~al.}(1978){Forman}, {Jones}, {Cominsky}, {Julien},
  {Murray}, {Peters}, {Tananbaum}, \& {Giacconi}}]{1978ApJS...38..357F}
{Forman}, W., {Jones}, C., {Cominsky}, L., {et~al.} 1978, \apjs, 38, 357

\bibitem[{{G{\'o}rski} {et~al.}(2005){G{\'o}rski}, {Hivon}, {Banday},
  {Wandelt}, {Hansen}, {Reinecke}, \& {Bartelmann}}]{2005ApJ...622..759G}
{G{\'o}rski}, K.~M., {Hivon}, E., {Banday}, A.~J., {et~al.} 2005, \apj, 622,
  759

\bibitem[{{Green}(2014)}]{2014BASI...42...47G}
{Green}, D.~A. 2014, Bulletin of the Astronomical Society of India, 42, 47

\bibitem[{{Hasinger} \& {van der Klis}(1989)}]{1989A&A...225...79H}
{Hasinger}, G., \& {van der Klis}, M. 1989, \aap, 225, 79

\bibitem[{{Hiroi} {et~al.}(2011){Hiroi}, {Ueda}, {Isobe}, {Hayashida},
  {Eguchi}, {Sugizaki}, {Kawai}, {Tsunemi}, {Matsuoka}, {Mihara}, {Yamaoka},
  {Ishikawa}, {Kimura}, {Kitayama}, {Kohama}, {Matsumura}, {Morii}, {Nakagawa},
  {Nakahira}, {Nakajima}, {Negoro}, {Serino}, {Shidatsu}, {Sootome},
  {Sugimori}, {Suwa}, {Toizumi}, {Tomida}, {Tsuboi}, {Ueno}, {Usui},
  {Yamamoto}, {Yamazaki}, \& {Yoshida}}]{2011PASJ...63S.677H}
{Hiroi}, K., {Ueda}, Y., {Isobe}, N., {et~al.} 2011, \pasj, 63, 677

\bibitem[{{Hiroi} {et~al.}(2013){Hiroi}, {Ueda}, {Hayashida}, {Shidatsu},
  {Sato}, {Kawamuro}, {Sugizaki}, {Nakahira}, {Serino}, {Kawai}, {Matsuoka},
  {Mihara}, {Morii}, {Nakajima}, {Negoro}, {Sakamoto}, {Tomida}, {Tsuboi},
  {Tsunemi}, {Ueno}, {Yamaoka}, {Yoshida}, {Asada}, {Eguchi}, {Hanayama},
  {Higa}, {Ishikawa}, {Ishikawa}, {Isobe}, {Kohama}, {Kimura}, {Morihana},
  {Nakagawa}, {Nakano}, {Nishimura}, {Ogawa}, {Sasaki}, {Sugimoto}, {Takagi},
  {Usui}, {Yamamoto}, {Yamauchi}, \& {Yoshidome}}]{2013ApJS..207...36H}
{Hiroi}, K., {Ueda}, Y., {Hayashida}, M., {et~al.} 2013, \apjs, 207, 36

\bibitem[{{Kimura} {et~al.}(2013){Kimura}, {Tsunemi}, {Tomida}, {Sugizaki},
  {Ueno}, {Hanayama}, {Yoshidome}, \& {Sasaki}}]{2013PASJ...65...14K}
{Kimura}, M., {Tsunemi}, H., {Tomida}, H., {et~al.} 2013, \pasj, 65, 14

\bibitem[{{Krimm} {et~al.}(2012){Krimm}, {Barthelmy}, {Baumgartner},
  {Cummings}, {Fenimore}, {Gehrels}, {Markwardt}, {Palmer}, {Sakamoto},
  {Skinner}, {Stamatikos}, {Tueller}, \& {Ukwatta}}]{2012ATel.4139....1K}
{Krimm}, H.~A., {Barthelmy}, S.~D., {Baumgartner}, W., {et~al.} 2012, The
  Astronomer's Telegram, 4139, 1

\bibitem[{{Levine} {et~al.}(1984){Levine}, {Lang}, {Lewin}, {Primini},
  {Dobson}, {Doty}, {Hoffman}, {Howe}, {Scheepmaker}, {Wheaton}, {Matteson},
  {Baity}, {Gruber}, {Knight}, {Nolan}, {Pelling}, {Rothschild}, \&
  {Peterson}}]{1984ApJS...54..581L}
{Levine}, A.~M., {Lang}, F.~L., {Lewin}, W.~H.~G., {et~al.} 1984, \apjs, 54,
  581

\bibitem[{{Matsuoka} {et~al.}(2009){Matsuoka}, {Kawasaki}, {Ueno}, {Tomida},
  {Kohama}, {Suzuki}, {Adachi}, {Ishikawa}, {Mihara}, {Sugizaki}, {Isobe},
  {Nakagawa}, {Tsunemi}, {Miyata}, {Kawai}, {Kataoka}, {Morii}, {Yoshida},
  {Negoro}, {Nakajima}, {Ueda}, {Chujo}, {Yamaoka}, {Yamazaki}, {Nakahira},
  {You}, {Ishiwata}, {Miyoshi}, {Eguchi}, {Hiroi}, {Katayama}, \&
  {Ebisawa}}]{2009PASJ...61..999M}
{Matsuoka}, M., {Kawasaki}, K., {Ueno}, S., {et~al.} 2009, \pasj, 61, 999

\bibitem[{{McHardy} {et~al.}(1981){McHardy}, {Lawrence}, {Pye}, \&
  {Pounds}}]{1981MNRAS.197..893M}
{McHardy}, I.~M., {Lawrence}, A., {Pye}, J.~P., \& {Pounds}, K.~A. 1981,
  \mnras, 197, 893

\bibitem[{{Mihara} {et~al.}(2011){Mihara}, {Nakajima}, {Sugizaki}, {Serino},
  {Matsuoka}, {Kohama}, {Kawasaki}, {Tomida}, {Ueno}, {Kawai}, {Kataoka},
  {Morii}, {Yoshida}, {Yamaoka}, {Nakahira}, {Negoro}, {Isobe}, {Yamauchi}, \&
  {Sakurai}}]{2011PASJ...63S.623M}
{Mihara}, T., {Nakajima}, M., {Sugizaki}, M., {et~al.} 2011, \pasj, 63, 623

\bibitem[{{Negoro} {et~al.}(2011){Negoro}, {Nakajima}, {Nakahira}, {Morii},
  {Krimm}, {Palmer}, {Kennea}, {Mihara}, {Sugizaki}, {Serino}, {Yamamoto},
  {Sootome}, {Matsuoka}, {Ueno}, {Tomida}, {Kohama}, {Ishikawa}, {Kawai},
  {Sugimori}, {Usui}, {Toizumi}, {Yoshida}, {Yamaoka}, {Tsunemi}, {Kimura},
  {Kitayama}, {Suwa}, {Asada}, {Sakakibara}, {Ueda}, {Hiroi}, {Shidatsu},
  {Tsuboi}, {Matsumura}, \& {Yamazaki}}]{2011ATel.3611....1N}
{Negoro}, H., {Nakajima}, M., {Nakahira}, S., {et~al.} 2011, The Astronomer's
  Telegram, 3611, 1

\bibitem[{{Nugent} {et~al.}(1983){Nugent}, {Jensen}, {Nousek}, {Garmire},
  {Mason}, {Walter}, {Bowyer}, {Stern}, \& {Riegler}}]{1983ApJS...51....1N}
{Nugent}, J.~J., {Jensen}, K.~A., {Nousek}, J.~A., {et~al.} 1983, \apjs, 51, 1

\bibitem[{{Palmer} {et~al.}(2005){Palmer}, {Barthelmey}, {Cummings}, {Gehrels},
  {Krimm}, {Markwardt}, {Sakamoto}, \& {Tueller}}]{2005ATel..546....1P}
{Palmer}, D.~M., {Barthelmey}, S.~D., {Cummings}, J.~R., {et~al.} 2005, The
  Astronomer's Telegram, 546, 1

\bibitem[{{Sugizaki} {et~al.}(2011){Sugizaki}, {Mihara}, {Serino}, {Yamamoto},
  {Matsuoka}, {Kohama}, {Tomida}, {Ueno}, {Kawai}, {Morii}, {Sugimori},
  {Nakahira}, {Yamaoka}, {Yoshida}, {Nakajima}, {Negoro}, {Eguchi}, {Isobe},
  {Ueda}, \& {Tsunemi}}]{2011PASJ...63S.635S}
{Sugizaki}, M., {Mihara}, T., {Serino}, M., {et~al.} 2011, \pasj, 63, 635

\bibitem[{{Sugizaki} {et~al.}(2013){Sugizaki}, {Yamaoka}, {Matsuoka}, {Kennea},
  {Mihara}, {Hiroi}, {Ishikawa}, {Isobe}, {Kawai}, {Kimura}, {Kitayama},
  {Kohama}, {Matsumura}, {Morii}, {Nakagawa}, {Nakahira}, {Nakajima}, {Negoro},
  {Serino}, {Shidatsu}, {Sootome}, {Sugimori}, {Suwa}, {Toizumi}, {Tomida},
  {Tsuboi}, {Tsunemi}, {Ueda}, {Ueno}, {Usui}, {Yamamoto}, {Yamauchi},
  {Yamazaki}, \& {Yoshida}}]{2013PASJ...65...58S}
{Sugizaki}, M., {Yamaoka}, K., {Matsuoka}, M., {et~al.} 2013, \pasj, 65, 58

\bibitem[{{Tawa} {et~al.}(2008){Tawa}, {Hayashida}, {Nagai}, {Nakamoto},
  {Tsunemi}, {Yamaguchi}, {Ishisaki}, {Miller}, {Mizuno}, {Dotani}, {Ozaki}, \&
  {Katayama}}]{2008PASJ...60S..11T}
{Tawa}, N., {Hayashida}, K., {Nagai}, M., {et~al.} 2008, \pasj, 60, 11

\bibitem[{{Tomida} {et~al.}(2011){Tomida}, {Tsunemi}, {Kimura}, {Kitayama},
  {Matsuoka}, {Ueno}, {Kawasaki}, {Katayama}, {Miyaguchi}, {Maeda}, {Daikyuji},
  \& {Isobe}}]{2011PASJ...63..397T}
{Tomida}, H., {Tsunemi}, H., {Kimura}, M., {et~al.} 2011, \pasj, 63, 397

\bibitem[{{Tsunemi} {et~al.}(2010){Tsunemi}, {Tomida}, {Katayama}, {Kimura},
  {Daikyuji}, {Miyaguchi}, {Maeda}, \& {MAXI Team}}]{2010PASJ...62.1371T}
{Tsunemi}, H., {Tomida}, H., {Katayama}, H., {et~al.} 2010, \pasj, 62, 1371

\bibitem[{{Usui} {et~al.}(2012){Usui}, {Nakahira}, {Tomida}, {Negoro}, {Morii},
  {Kawai}, {Ishikawa}, {Ueno}, {Ishikawa}, {Mihara}, {Sugizaki}, {Serino},
  {Yamamoto}, {Matsuoka}, {Yoshida}, {Tsunemi}, {Kimura}, {Nakajima}, {Asada},
  {Sakakibara}, {Serita}, {Ueda}, {Hiroi}, {Shidatsu}, {Sato}, {Tsuboi},
  {Yamauchi}, {Nishimura}, {Hanayama}, {Yoshidome}, \&
  {Yamaoka}}]{2012ATel.4140....1U}
{Usui}, R., {Nakahira}, S., {Tomida}, H., {et~al.} 2012, The Astronomer's
  Telegram, 4140, 1

\bibitem[{{Vanderspek} {et~al.}(2005){Vanderspek}, {Morgan}, {Crew},
  {Graziani}, \& {Suzuki}}]{2005ATel..516....1V}
{Vanderspek}, R., {Morgan}, E., {Crew}, G., {Graziani}, C., \& {Suzuki}, M.
  2005, The Astronomer's Telegram, 516, 1

\bibitem[{{Voges} {et~al.}(1999){Voges}, {Aschenbach}, {Boller},
  {Br{\"a}uninger}, {Briel}, {Burkert}, {Dennerl}, {Englhauser}, {Gruber},
  {Haberl}, {Hartner}, {Hasinger}, {K{\"u}rster}, {Pfeffermann}, {Pietsch},
  {Predehl}, {Rosso}, {Schmitt}, {Tr{\"u}mper}, \&
  {Zimmermann}}]{1999A&A...349..389V}
{Voges}, W., {Aschenbach}, B., {Boller}, T., {et~al.} 1999, \aap, 349, 389

\bibitem[{{Voges} {et~al.}(2000){Voges}, {Aschenbach}, {Boller}, {Brauninger},
  {Briel}, {Burkert}, {Dennerl}, {Englhauser}, {Gruber}, {Haberl}, {Hartner},
  {Hasinger}, {Pfeffermann}, {Pietsch}, {Predehl}, {Schmitt}, {Trumper}, \&
  {Zimmermann}}]{2000yCat.9029....0V}
---. 2000, VizieR Online Data Catalog, 9029, 0

\bibitem[{{Warwick} {et~al.}(1981){Warwick}, {Marshall}, {Fraser}, {Watson},
  {Lawrence}, {Page}, {Pounds}, {Ricketts}, {Sims}, \&
  {Smith}}]{1981MNRAS.197..865W}
{Warwick}, R.~S., {Marshall}, N., {Fraser}, G.~W., {et~al.} 1981, \mnras, 197,
  865

\bibitem[{{Wood} {et~al.}(1984){Wood}, {Meekins}, {Yentis}, {Smathers},
  {McNutt}, {Bleach}, {Friedman}, {Byram}, {Chubb}, \&
  {Meidav}}]{1984ApJS...56..507W}
{Wood}, K.~S., {Meekins}, J.~F., {Yentis}, D.~J., {et~al.} 1984, \apjs, 56, 507

\bibitem[{{Yan} {et~al.}(2014){Yan}, {Zhang}, {Zhang}, {Stiele}, \&
  {Yu}}]{2014ATel.6649....1Y}
{Yan}, Z., {Zhang}, W., {Zhang}, H., {Stiele}, H., \& {Yu}, W. 2014, The
  Astronomer's Telegram, 6649, 1

\end{thebibliography}


\clearpage
\newpage

\setlength{\headsep}{7.0cm}
\begin{landscape}
\begin{longtable}{cccccccclc} \hline
 \caption{the MXAI/SSC catalog.
   The flux is calculated only when S$_D >$5.0 in the energy band.}
 \label{ssc_catalog}
 \hline
   \multicolumn{1}{c}{(1)} & 
   \multicolumn{1}{c}{(2)} & 
   \multicolumn{1}{c}{(3)} & 
   \multicolumn{1}{c}{(4)} & 
   \multicolumn{1}{c}{(5)} & 
   \multicolumn{1}{c}{(6)} & 
   \multicolumn{1}{c}{(7)} &
   \multicolumn{1}{c}{(8)} & 
   \multicolumn{1}{c}{(9)} &
   \multicolumn{1}{c}{(10)} \\ 
   \multicolumn{1}{c}{No.} & 
   \multicolumn{1}{c}{MAXI SSC Name} & 
   \multicolumn{1}{c}{RA}  & 
   \multicolumn{1}{c}{DEC} & 
   \multicolumn{2}{c}{\underline{\ \ \ \ \ \ \ $S_D$\ \ \ \ \ \ \ }} & 
   \multicolumn{2}{c}{\underline{\ \ \ \ \ \ \ flux\ \ \ \ \ \ \ }} & 
   \multicolumn{1}{c}{identification} &
   \multicolumn{1}{c}{GSC-ID} \\ 
   \multicolumn{1}{c}{} & 
   \multicolumn{1}{c}{} & 
   \multicolumn{1}{c}{degree}  & 
   \multicolumn{1}{c}{degree}  & 
   \multicolumn{1}{c}{} & 
   \multicolumn{1}{c}{} & 
   \multicolumn{2}{c}{\underline{10$^{-11}$erg sec$^{-1}$ cm$^{-2}$}}   & 
   \multicolumn{1}{c}{} &
   \multicolumn{1}{c}{} \\ 
   \multicolumn{1}{c}{} & 
   \multicolumn{1}{c}{} & 
   \multicolumn{1}{c}{}  & 
   \multicolumn{1}{c}{}  & 
   \multicolumn{1}{c}{soft} & 
   \multicolumn{1}{c}{hard} & 
   \multicolumn{1}{c}{soft}   & 
   \multicolumn{1}{c}{hard}   & 
   \multicolumn{1}{c}{} &
   \multicolumn{1}{c}{} \\ \hline
 \hline
 \endfirsthead
 \hline
   \multicolumn{1}{c}{(1)} & 
   \multicolumn{1}{c}{(2)} & 
   \multicolumn{1}{c}{(3)} & 
   \multicolumn{1}{c}{(4)} & 
   \multicolumn{1}{c}{(5)} & 
   \multicolumn{1}{c}{(6)} & 
   \multicolumn{1}{c}{(7)} &
   \multicolumn{1}{c}{(8)} & 
   \multicolumn{1}{c}{(9)} &
   \multicolumn{1}{c}{(10)} \\ 
   \multicolumn{1}{c}{No.} & 
   \multicolumn{1}{c}{MAXI SSC Name} & 
   \multicolumn{1}{c}{RA}  & 
   \multicolumn{1}{c}{DEC} & 
   \multicolumn{2}{c}{\underline{\ \ \ \ \ \ \ $S_D$\ \ \ \ \ \ \ }} & 
   \multicolumn{2}{c}{\underline{\ \ \ \ \ \ \ flux\ \ \ \ \ \ \ }} & 
   \multicolumn{1}{c}{identification} &
   \multicolumn{1}{c}{GSC ID} \\ 
   \multicolumn{1}{c}{} & 
   \multicolumn{1}{c}{} & 
   \multicolumn{1}{c}{degree}  & 
   \multicolumn{1}{c}{degree}  & 
   \multicolumn{1}{c}{} & 
   \multicolumn{1}{c}{} & 
   \multicolumn{2}{c}{\underline{10$^{-11}$erg sec$^{-1}$ cm$^{-2}$}}   & 
   \multicolumn{1}{c}{} &
   \multicolumn{1}{c}{} \\ 
   \multicolumn{1}{c}{} & 
   \multicolumn{1}{c}{} & 
   \multicolumn{1}{c}{}  & 
   \multicolumn{1}{c}{}  & 
   \multicolumn{1}{c}{soft} & 
   \multicolumn{1}{c}{hard} & 
   \multicolumn{1}{c}{soft}   & 
   \multicolumn{1}{c}{hard}   & 
   \multicolumn{1}{c}{} &
   \multicolumn{1}{c}{} \\ \hline
 \hline
 \endhead
 \hline
 \endfoot
 \hline
 \endlastfoot
1 & 1MAXIS~J0006+728 & 1.73 & 72.87 & 6.0 & 1.7 & 3.3 & - & CTA 1(SNR) & 2MAXI J0004+726 \\
2 & 1MAXIS~J0025+641 & 6.41 & 64.14 & 28.4 & 22.9 & 22.9 & 31.1 & Tycho SNR(SNR) &  \\
3 & 1MAXIS~J0034+596 & 8.61 & 59.65 & 4.3 & 5.9 & - & 8.6 & 1ES 0033+595(AGN) &  \\
4 & 1MAXIS~J0042-094 & 10.52 & -9.45 & 8.1 & 4.9 & 3.7 & - & Abell 85(GC) & 2MAXI J0041-092 \\
5 & 1MAXIS~J0056+606 & 14.04 & 60.68 & 8.7 & 9.7 & 5.1 & 13.0 & Gamma Cas(Star) &  \\
6 & 1MAXIS~J0116-735 & 19.16 & -73.52 & 5.9 & 12.3 & 2.3 & 13.0 & SMC X-1(Binary-Pulsar) & 2MAXI J0117-734 \\
7 & 1MAXIS~J0146+617 & 26.65 & 61.75 & 10.7 & 6.8 & 6.8 & 11.4 & 4U 0142+61(Isolated Pulser) &  \\
8 & 1MAXIS~J0153+361 & 28.31 & 36.14 & 5.3 & 3.0 & 3.0 & - & Abell 262(GC) & 2MAXI J0151+361 \\
9 & 1MAXIS~J0246-584 & 41.61 & -58.47 & 6.8 & 0.4 & 2.5 & - & 2MAXI J0243-582(Unknown) & 2MAXI J0243-582 \\
10 & 1MAXIS~J0254+415 & 43.74 & 41.55 & 7.2 & 6.7 & 4.7 & 10.1 & AWM 7(GC) & 2MAXI J0254+416 \\
11 & 1MAXIS~J0258+133 & 44.65 & 13.35 & 7.2 & 10.2 & 2.8 & 10.7 & Abell 401(GC) & 2MAXI J0258+135 \\
12 & 1MAXIS~J0320+415 & 50.01 & 41.52 & 42.4 & 49.5 & 37.0 & 94.3 & Perseus Cluster(GC) & 2MAXI J0319+415 \\
13 & 1MAXIS~J0326+286 & 51.59 & 28.62 & 9.0 & 3.6 & 2.7 & - & UX Ari(Star) & 2MAXI J0326+287 \\
14 & 1MAXIS~J0337+006 & 54.28 & 0.66 & 10.0 & 5.2 & 5.0 & 6.9 & HR1099(Star) & 2MAXI J0336+006 \\
15 & 1MAXIS~J0338+100 & 54.63 & 10.04 & 6.1 & 4.4 & 3.3 & - & ZwCl 0335+0956(GC) & 2MAXI J0339+100 \\
16 & 1MAXIS~J0339-353 & 54.83 & -35.39 & 9.5 & 3.6 & 5.8 & - & FORNAX(GC) &  \\
17 & 1MAXIS~J0354+310 & 58.72 & 31.05 & 15.9 & 38.0 & 8.5 & 50.7 & X Per(Binary-Pulsar) & 2MAXI J0355+311 \\
18 & 1MAXIS~J0413+105 & 63.44 & 10.54 & 6.6 & 7.5 & 2.6 & 6.9 & Abell 478(GC) & 2MAXI J0413+106 \\
19 & 1MAXIS~J0430-613 & 67.61 & -61.36 & 5.8 & 4.2 & 3.5 & - & Abell 3266(GC) & 2MAXI J0431-613 \\
20 & 1MAXIS~J0433-132 & 68.39 & -13.27 & 8.1 & 4.9 & 3.4 & - & Abell 496(GC) & 2MAXI J0433-131 \\
21 & 1MAXIS~J0450+450 & 72.63 & 45.02 & 7.0 & 10.8 & 4.8 & 15.5 & 3C 129(AGN) &  \\
22 & 1MAXIS~J0451-037 & 72.91 & -3.75 & 2.2 & 5.9 & - & 2.4 & MCG -01-13-025(AGN) & 2MAXI J0450-035 \\
23 & 1MAXIS~J0500+518 & 75.10 & 51.82 & 16.7 & 1.3 & 12.5 & - & G156.2+5.7(SNR) &  \\
24 & 1MAXIS~J0500+461 & 75.24 & 46.17 & 8.1 & -0.6 & 6.3 & - & HB9(SNR) &  \\
25 & 1MAXIS~J0514-401 & 78.60 & -40.15 & 9.0 & 9.7 & 6.7 & 11.5 & 4U 0513-40(Binary-NS) & 2MAXI J0514-399 \\
26 & 1MAXIS~J0517+459 & 79.37 & 45.97 & 9.1 & -2.1 & 6.0 & - & 1RXS J051642.2+460001(GC) &  \\
27 & 1MAXIS~J0518-720 & 79.66 & -72.02 & 17.3 & 27.1 & 9.6 & 36.4 & LMC X-2(Binary-NS) & 2MAXI J0520-719 \\
28 & 1MAXIS~J0531-661 & 82.98 & -66.12 & 15.5 & 8.5 & 8.8 & 8.7 & LMC X-4(Binary-Pulsar) & 2MAXI J0532-663 \\
29 & 1MAXIS~J0534+220 & 83.64 & 22.03 & 309.5 & 366.4 & 816.2 & 1816.0 & Crab(Isolated-Pulsar) &  \\
30 & 1MAXIS~J0535-054 & 83.78 & -5.40 & 15.9 & 10.6 & 6.8 & 11.0 & 2MAXI J0535-052(Star Culster) & 2MAXI J0535-052 \\
31 & 1MAXIS~J0536-697 & 84.20 & -69.72 & 36.4 & 31.0 & 30.7 & 46.9 & LMC X-1(Binary-BH) & 2MAXI J0539-696 \\
32 & 1MAXIS~J0538-641 & 84.65 & -64.18 & 25.0 & 21.9 & 17.2 & 28.2 & LMC X-3(Binary-BH) & 2MAXI J0539-640 \\
33 & 1MAXIS~J0539+263 & 84.81 & 26.31 & 7.5 & 25.7 & 3.5 & 31.3 & A 0535+262(Binary-Pulsar) &  \\
34 & 1MAXIS~J0551-074 & 87.97 & -7.48 & 0.5 & 5.7 & - & 8.3 & NGC 2110(AGN) & 2MAXI J0552-073 \\
35 & 1MAXIS~J0556-332 & 89.18 & -33.22 & 7.2 & 8.3 & 4.0 & 9.7 & MAXI J0556-332(Binary-NS) & 2MAXI J0556-331 \\
36 & 1MAXIS~J0602+286 & 90.65 & 28.69 & 5.9 & 5.6 & 2.9 & 6.4 & MAXI J0602+291(Unknown) &  \\
37 & 1MAXIS~J0603+292 & 90.79 & 29.23 & 5.2 & 6.7 & 2.1 & 5.7 & MAXI J0602+291(Unknown) &  \\
38 & 1MAXIS~J0617+091 & 94.30 & 9.17 & 69.9 & 68.9 & 55.8 & 108.3 & H 0614+091(Binary-NS) &  \\
39 & 1MAXIS~J0617+225 & 94.30 & 22.59 & 39.9 & 9.1 & 21.7 & 10.0 & IC 443 SNR(SNR) &  \\
40 & 1MAXIS~J0743+288 & 115.93 & 28.84 & 6.8 & 4.0 & 2.9 & - & 2MAXI J0744+288(Unknown) & 2MAXI J0744+288 \\
41 & 1MAXIS~J0817-075 & 124.40 & -7.58 & 6.7 & 5.4 & 2.5 & 5.8 & Abell 644(GC) & 2MAXI J0817-074 \\
42 & 1MAXIS~J0822-429 & 125.71 & -42.94 & 152.1 & 19.1 & 339.3 & 29.6 & Pup~A(SNR) &  \\
43 & 1MAXIS~J0834-455 & 128.65 & -45.54 & 19.4 & 15.6 & 9.6 & 22.2 & Vela Pulsar/SNR(Isolated Pulsar/SNR) &  \\
44 & 1MAXIS~J0850-461 & 132.56 & -46.19 & 27.6 & 7.0 & 26.5 & 7.1 & RX J0852.0-4622(SNR) &  \\
45 & 1MAXIS~J0902-406 & 135.51 & -40.60 & 4.0 & 39.0 & - & 65.1 & Vela X-1(Binary-Pulsar) &  \\
46 & 1MAXIS~J0908-097 & 137.23 & -9.77 & 7.4 & 7.5 & 3.6 & 9.0 & Abell 754(GC) & 2MAXI J0909-096 \\
47 & 1MAXIS~J0917-120 & 139.44 & -12.06 & 6.3 & 5.2 & 2.4 & 4.5 & MAXI J0918-121(Unknown) & 2MAXI J0918-119 \\
48 & 1MAXIS~J0920-553 & 140.09 & -55.30 & 13.4 & 15.9 & 10.1 & 23.7 & 1H 0918-548(Binary-NS) &  \\
49 & 1MAXIS~J0922-631 & 140.70 & -63.14 & 3.4 & 5.5 & - & 7.0 & 2S 0921-63(Binary-NS) &  \\
50 & 1MAXIS~J0924-317 & 141.15 & -31.76 & 4.3 & 7.2 & - & 8.8 & X 0922-314(Unknown) & 2MAXI J0924-316 \\
51 & 1MAXIS~J0947-310 & 146.83 & -31.04 & 0.9 & 5.2 & - & 8.7 & MCG -05-23-016(AGN) & 2MAXI J0947-308 \\
52 & 1MAXIS~J1009-583 & 152.38 & -58.36 & 3.0 & 24.0 & - & 38.2 & GRO J1008-57(Binary-Pulsar) &  \\
53 & 1MAXIS~J1020-038 & 155.12 & -3.81 & 2.3 & 5.6 & - & 1.4 & 2MAXI J1020-034(AGN) & 2MAXI J1020-034 \\
54 & 1MAXIS~J1029-353 & 157.39 & -35.40 & 6.1 & 3.1 & 4.5 & - & 2MAXI J1030-351(Unknown) & 2MAXI J1030-351 \\
55 & 1MAXIS~J1036-274 & 159.24 & -27.43 & 7.5 & 5.3 & 3.3 & 7.6 & Abell 1060(GC) & 2MAXI J1036-275 \\
56 & 1MAXIS~J1045-597 & 161.26 & -59.76 & 15.6 & 9.7 & 11.6 & 12.6 & Eta Car(Star) &  \\
57 & 1MAXIS~J1102-610 & 165.50 & -61.05 & 6.2 & -8.9 & 5.2 & - & MSH 11-61A(SNR) &  \\
58 & 1MAXIS~J1104+381 & 166.11 & 38.16 & 29.0 & 15.2 & 16.1 & 17.1 & Mrk 421(AGN) & 2MAXI J1104+382 \\
59 & 1MAXIS~J1120-606 & 170.15 & -60.68 & 14.6 & 52.1 & 7.8 & 85.5 & Cen X-3(Binary-Pulsar) &  \\
60 & 1MAXIS~J1123-593 & 170.96 & -59.37 & 22.4 & 21.8 & 13.0 & 6.4 & MSH 11-54(SNR) &  \\
61 & 1MAXIS~J1139-654 & 175.00 & -65.45 & 9.6 & 5.8 & 4.6 & 7.4 & 4U 1137-65(Star) &  \\
62 & 1MAXIS~J1144+197 & 176.16 & 19.77 & 10.3 & 9.0 & 4.1 & 7.2 & Abell 1367(GC) & 2MAXI J1144+198 \\
63 & 1MAXIS~J1147-623 & 176.99 & -62.32 & 6.1 & 7.2 & 3.7 & 7.0 & 1E1145.1-6141(Binary-Pulsar) &  \\
64 & 1MAXIS~J1210+392 & 182.64 & 39.22 & 0.6 & 9.3 & - & 14.2 & NGC 4151(AGN) & 2MAXI J1210+394 \\
65 & 1MAXIS~J1216-652 & 184.06 & -65.26 & 7.5 & -0.6 & 4.5 & - & 4U 1210-64(Binary-NS) &  \\
66 & 1MAXIS~J1219+301 & 184.95 & 30.17 & 11.5 & 6.9 & 4.7 & 6.9 & Mrk 766(AGN) & 2MAXI J1220+300 \\
67 & 1MAXIS~J1225-629 & 186.43 & -62.91 & 2.1 & 15.8 & - & 23.5 & GX 301-2(Binary-Pulsar) &  \\
68 & 1MAXIS~J1229+020 & 187.27 & 2.09 & 7.2 & 9.8 & 3.7 & 8.3 & 3C 273(AGN) & 2MAXI J1229+021 \\
69 & 1MAXIS~J1229+082 & 187.50 & 8.26 & 6.4 & 4.8 & 2.3 & - & 1H1228+081(Unknown) &  \\
70 & 1MAXIS~J1230+123 & 187.69 & 12.37 & 57.2 & 39.0 & 39.3 & 46.2 & M 87(AGN) & 2MAXI J1230+124 \\
71 & 1MAXIS~J1249-413 & 192.27 & -41.32 & 14.2 & 13.6 & 11.6 & 18.4 & Centaurus Cluster(GC) & 2MAXI J1249-412 \\
72 & 1MAXIS~J1250-592 & 192.61 & -59.25 & 7.8 & 9.3 & 5.2 & 10.7 & 1A 1246-588(Binary-NS) &  \\
73 & 1MAXIS~J1253-295 & 193.45 & -29.52 & 8.7 & 12.4 & 5.6 & 12.0 & EX Hya(Binary-WD) & 2MAXI J1252-291 \\
74 & 1MAXIS~J1257-693 & 194.27 & -69.35 & 25.5 & 40.9 & 18.0 & 63.0 & 4U 1254-690(Binary-NS) &  \\
75 & 1MAXIS~J1259+278 & 194.87 & 27.89 & 28.2 & 28.7 & 14.4 & 32.7 & Coma Cluster(GC) & 2MAXI J1259+279 \\
76 & 1MAXIS~J1300-617 & 195.18 & -61.72 & 0.3 & 23.2 & - & 36.7 & GX 304-1(Binary-Pulsar) &  \\
77 & 1MAXIS~J1316-163 & 199.09 & -16.33 & 5.9 & 2.8 & 2.8 & - & NGC 5044(GC) &  \\
78 & 1MAXIS~J1325-431 & 201.39 & -43.15 & -1.5 & 11.5 & - & 15.9 & Cen A(AGN) & 2MAXI J1325-429 \\
79 & 1MAXIS~J1325-626 & 201.49 & -62.62 & 0.0 & 7.3 & - & 7.3 & 4U 1323-619(Binary-NS) &  \\
80 & 1MAXIS~J1329-316 & 202.48 & -31.66 & 8.3 & 4.3 & 5.9 & - & Abell 3558(GC) & 2MAXI J1329-315 \\
81 & 1MAXIS~J1347-329 & 206.95 & -32.90 & 4.9 & 7.7 & - & 7.9 & Abell 3571(GC) & 2MAXI J1347-328 \\
82 & 1MAXIS~J1349+265 & 207.26 & 26.56 & 7.0 & 4.1 & 3.1 & - & Abell 1795(GC) & 2MAXI J1348+267 \\
83 & 1MAXIS~J1349-303 & 207.32 & -30.39 & 4.0 & 6.2 & - & 7.8 & IC 4329A(AGN) & 2MAXI J1349-302 \\
84 & 1MAXIS~J1413-033 & 213.46 & -3.31 & 0.0 & 8.4 & - & 7.6 & NGC 5506(AGN) & 2MAXI J1413-030 \\
85 & 1MAXIS~J1441-627 & 220.44 & -62.73 & 17.6 & 3.0 & 14.0 & - & RCW 86(SNR) &  \\
86 & 1MAXIS~J1502-420 & 225.75 & -42.04 & 10.9 & 5.2 & 6.7 & 6.8 & SN 1006(SNR) & 2MAXI J1503-421 \\
87 & 1MAXIS~J1511+058 & 227.98 & 5.85 & 5.6 & 8.7 & 3.1 & 7.2 & Abell 2029(GC) & 2MAXI J1511+059 \\
88 & 1MAXIS~J1513-591 & 228.40 & -59.14 & 7.0 & 10.1 & 4.7 & 12.4 & PSR B1509-58(Isolated Pulsor) &  \\
89 & 1MAXIS~J1520-572 & 230.07 & -57.28 & 1.2 & 18.3 & - & 15.0 & Cir X-1(Binary-NS) &  \\
90 & 1MAXIS~J1541-525 & 235.49 & -52.52 & -0.8 & 12.6 & - & 15.0 & 4U 1538-52(Binary-Pulsar) &  \\
91 & 1MAXIS~J1544-563 & 236.05 & -56.39 & 1.2 & 8.7 & - & 10.5 & MAXI J1543-564(Binary-BH) &  \\
92 & 1MAXIS~J1547-626 & 236.82 & -62.67 & 25.8 & 36.0 & 22.6 & 59.2 & 4U 1543-624(Binary-NS/BH) &  \\
93 & 1MAXIS~J1558+270 & 239.64 & 27.09 & 5.3 & 6.3 & 2.3 & 7.6 & Abell 2142(GC) & 2MAXI J1558+272 \\
94 & 1MAXIS~J1603+162 & 240.77 & 16.22 & 7.9 & 6.3 & 3.9 & 5.0 & Abell 2147(GC) & 2MAXI J1602+161 \\
95 & 1MAXIS~J1603-608 & 241.00 & -60.82 & 10.4 & 19.7 & 2.5 & 13.9 & 1H 1556-605(Binary-NS/BH) &  \\
96 & 1MAXIS~J1610-608 & 242.69 & -60.81 & 14.1 & 19.2 & 5.4 & 11.8 & 2MASX J16115141-6037549(AGN) &  \\
97 & 1MAXIS~J1612-524 & 243.11 & -52.47 & 44.1 & 107.1 & 49.9 & 303.0 & 4U 1608-52(Binary-NS) &  \\
98 & 1MAXIS~J1614+338 & 243.67 & 33.89 & 8.4 & 1.2 & 5.2 & - & HB89 1611+343(AGN) &  \\
99 & 1MAXIS~J1619-156 & 244.98 & -15.63 & 801.6 & 1270.3 & 5258.0 & 20635.5 & Sco X-1(Binary-NS) & 2MAXI J1619-155 \\
100 & 1MAXIS~J1629+393 & 247.29 & 39.36 & 6.1 & 9.0 & 4.3 & 12.9 & Abell 2199(GC) & 2MAXI J1628+396 \\
101 & 1MAXIS~J1632-675 & 248.05 & -67.56 & 15.0 & 19.1 & 9.7 & 29.1 & 4U 1626-67(Binary-Pulsar) & 2MAXI J1632-674 \\
102 & 1MAXIS~J1638-644 & 249.64 & -64.43 & 9.2 & 13.6 & 5.1 & 15.2 & TrA Cluster(GC) & 2MAXI J1638-643 \\
103 & 1MAXIS~J1640-538 & 250.18 & -53.80 & 40.1 & 59.4 & 39.1 & 112.0 & H 1636-536(Binary-NS) &  \\
104 & 1MAXIS~J1645-456 & 251.41 & -45.67 & 11.6 & 160.3 & 9.6 & 648.5 & GX 340+0(Binary-NS) &  \\
105 & 1MAXIS~J1654+396 & 253.64 & 39.66 & 11.7 & 10.6 & 9.2 & 14.0 & Mrk 501(AGN) & 2MAXI J1653+398 \\
106 & 1MAXIS~J1657+352 & 254.48 & 35.29 & 6.4 & 11.0 & 3.1 & 14.8 & Her X-1(Binary-Pulsar) & 2MAXI J1658+353 \\
107 & 1MAXIS~J1705+787 & 256.42 & 78.79 & 7.5 & 9.7 & 2.7 & 6.7 & Abell 2256(GC) & 2MAXI J1704+785 \\
108 & 1MAXIS~J1705-364 & 256.44 & -36.46 & 116.4 & 267.9 & 184.9 & 1293.8 & GX 349+2(Binary-NS) &  \\
109 & 1MAXIS~J1706+240 & 256.59 & 24.01 & 2.5 & 5.8 & - & 6.6 & 4U 1700+24(Binary-NS) & 2MAXI J1706+240 \\
110 & 1MAXIS~J1708-440 & 257.13 & -44.09 & 36.9 & 127.3 & 39.9 & 390.0 & 4U 1705-440(Binary-NS) &  \\
111 & 1MAXIS~J1712-407 & 258.06 & -40.76 & 13.9 & 37.2 & 1.9 & 29.5 & 4U 1708-40(Binary-NS) &  \\
112 & 1MAXIS~J1712-234 & 258.09 & -23.41 & 22.2 & 34.9 & 14.9 & 48.1 & Ophiuchus Cluster(GC) &  \\
113 & 1MAXIS~J1713-399 & 258.32 & -39.94 & 16.8 & 33.2 & 7.2 & 11.0 & RX J1713.7-3946(SNR) &  \\
114 & 1MAXIS~J1727-308 & 261.90 & -30.83 & 12.3 & 27.7 & 8.1 & 41.1 & Terzan 2/4U 1722-30(Binary-NS) &  \\
115 & 1MAXIS~J1730-215 & 262.73 & -21.59 & 14.5 & 2.1 & 9.5 & - & Kepler SNR(SNR) &  \\
116 & 1MAXIS~J1731-169 & 262.93 & -16.96 & 93.9 & 148.6 & 105.2 & 377.9 & GX 9+9(Binary-NS) &  \\
117 & 1MAXIS~J1732-338 & 263.02 & -33.87 & 12.2 & 68.6 & 10.7 & 155.6 & GX 354-0(Binary-NS) &  \\
118 & 1MAXIS~J1738-444 & 264.70 & -44.49 & 60.4 & 102.6 & 80.7 & 279.6 & H 1735-444(Binary-NS) &  \\
119 & 1MAXIS~J1745-292 & 266.40 & -29.26 & 1.1 & 85.4 & - & 214.2 & KS 1741-293(Binary-NS) &  \\
120 & 1MAXIS~J1746-322 & 266.52 & -32.26 & 1.5 & 19.7 & - & 24.4 & H 1743-322(Binary-BH) &  \\
121 & 1MAXIS~J1747-267 & 266.94 & -26.78 & 44.5 & 177.7 & 47.2 & 639.8 & GX 3+1(Binary-NS) &  \\
122 & 1MAXIS~J1749-370 & 267.50 & -37.05 & 20.4 & 41.6 & 17.2 & 68.8 & 4U 1746-37(Binary-NS) &  \\
123 & 1MAXIS~J1753-014 & 268.36 & -1.43 & 55.9 & 38.8 & 38.7 & 54.8 & Swift J1753.5-0127(Binary-BH) & 2MAXI J1753-013 \\
124 & 1MAXIS~J1801-250 & 270.28 & -25.09 & 50.7 & 313.3 & 53.8 & 1759.5 & GX 5-1(Binary-NS) &  \\
125 & 1MAXIS~J1801-205 & 270.38 & -20.52 & 58.6 & 229.6 & 59.3 & 965.7 & GX 9+1(Binary-NS) &  \\
126 & 1MAXIS~J1811-229 & 272.87 & -22.98 & 5.6 & 7.6 & 4.0 & 14.5 & V5588 Sgr(Binary-WD) &  \\
127 & 1MAXIS~J1814-171 & 273.64 & -17.15 & 31.3 & 176.1 & 23.8 & 578.2 & GX 13+1(Binary-NS) &  \\
128 & 1MAXIS~J1816-140 & 274.01 & -14.03 & 58.6 & 250.6 & 53.7 & 1047.3 & GX 17+2(Binary-NS) &  \\
129 & 1MAXIS~J1823-303 & 275.90 & -30.37 & 95.2 & 143.6 & 150.7 & 461.8 & NGC 6624(Binary-NS) &  \\
130 & 1MAXIS~J18250000 & 276.36 & -0.01 & 14.8 & 31.5 & 8.8 & 48.5 & H 1822-000(Binary-NS/BH) &  \\
131 & 1MAXIS~J1825-371 & 276.37 & -37.15 & 2.9 & 18.6 & - & 30.9 & 4U 1822-371 V691 CrA(Binary-NS) & 2MAXI J1825-370 \\
132 & 1MAXIS~J1829-238 & 277.37 & -23.83 & 17.2 & 40.2 & 12.4 & 62.6 & GS 1826-238(Binary-NS) &  \\
133 & 1MAXIS~J1832-103 & 278.09 & -10.31 & -1.3 & 9.6 & - & 14.3 & G21.5-0.9(SNR) &  \\
134 & 1MAXIS~J1835-193 & 278.94 & -19.31 & 9.8 & 5.8 & 4.7 & 5.7 & MAXI J1836-194(Binary-BH) &  \\
135 & 1MAXIS~J1840+050 & 280.02 & 5.06 & 76.7 & 125.0 & 96.4 & 377.6 & Ser X-1(Binary-NS) &  \\
136 & 1MAXIS~J1841-053 & 280.33 & -5.30 & 7.3 & 13.0 & 1.1 & 10.7 & 1E 1841-045(Isolated Pulsar) &  \\
137 & 1MAXIS~J1844-058 & 281.03 & -5.81 & 8.9 & 7.8 & 2.0 & 3.1 &  &  \\
138 & 1MAXIS~J1846-029 & 281.59 & -2.97 & 0.8 & 11.3 & - & 10.9 & AX J1844.8-0258/AX J1845-0258(Binary-NS/BH) &  \\
139 & 1MAXIS~J1852+797 & 283.09 & 79.79 & 5.4 & 9.4 & 1.2 & 4.1 & 3C 390.3(AGN) & 2MAXI J1842+797 \\
140 & 1MAXIS~J1853-086 & 283.27 & -8.69 & 12.3 & 11.2 & 7.5 & 15.8 & 4U 1850-086(Binary-NS) &  \\
141 & 1MAXIS~J1855-009 & 283.86 & -0.98 & 5.0 & 1.5 & 4.6 & - &  &  \\
142 & 1MAXIS~J1855+013 & 283.99 & 1.32 & 9.5 & 8.0 & 5.8 & 9.7 & 1H1852+015(Unknown) &  \\
143 & 1MAXIS~J1900-249 & 285.03 & -24.95 & 26.3 & 27.5 & 23.3 & 44.1 & HETE J1900.1-2455(Binary-NS) & 2MAXI J1900-248 \\
144 & 1MAXIS~J1910-057 & 287.61 & -5.78 & 114.9 & 80.9 & 218.3 & 219.5 & MAXI J1910-057(Binary-BH) &  \\
145 & 1MAXIS~J1911+005 & 287.81 & 0.50 & 12.8 & 17.0 & 9.6 & 28.4 & Aql X-1(Binary-NS) &  \\
146 & 1MAXIS~J1912+048 & 288.05 & 4.88 & 5.2 & 8.1 & 3.3 & 9.6 & SS 433(Binary-NS/BH) &  \\
147 & 1MAXIS~J1915+110 & 288.84 & 11.03 & 37.3 & 257.6 & 28.3 & 1294.7 & GRS 1915+105(Binary-BH) &  \\
148 & 1MAXIS~J1918-051 & 289.74 & -5.18 & 6.6 & 12.2 & 0.9 & 15.6 & 4U 1916-053(Binary-NS) &  \\
149 & 1MAXIS~J1921+438 & 290.35 & 43.87 & 10.0 & 9.4 & 6.9 & 10.4 & Abell 2319(GC) & 2MAXI J1921+440 \\
150 & 1MAXIS~J1930+097 & 292.74 & 9.79 & 5.3 & 6.1 & 1.9 & -6.1 & 2FGL J1931.1+0938(AGN) &  \\
151 & 1MAXIS~J1933+312 & 293.32 & 31.25 & 7.8 & 0.5 & 4.3 & - & G65.3+5.7(SNR) &  \\
152 & 1MAXIS~J1937-062 & 294.40 & -6.22 & 5.0 & 1.5 & 2.5 & - & 1H 1934-063(AGN) & 2MAXI J1937-060 \\
153 & 1MAXIS~J1958+351 & 299.58 & 35.14 & 306.3 & 274.7 & 1054.8 & 1348.8 & Cyg X-1(Binary-BH) &  \\
154 & 1MAXIS~J1959+116 & 299.86 & 11.68 & 38.0 & 51.1 & 30.1 & 96.8 & 4U 1957+115(Binary-BH) &  \\
155 & 1MAXIS~J2001+651 & 300.29 & 65.16 & 12.4 & 7.7 & 6.6 & 8.0 & 1ES 1959+650(AGN) & 2MAXI J1959+651 \\
156 & 1MAXIS~J2019+404 & 304.89 & 40.41 & 7.8 & 1.1 & 5.0 & - & Cygnus Super Bubble(SNR) &  \\
157 & 1MAXIS~J2032+374 & 308.07 & 37.48 & -2.1 & 11.8 & - & 14.5 & EXO 2030+375(Binary-Pulsar) &  \\
158 & 1MAXIS~J2032+409 & 308.09 & 40.91 & 2.6 & 87.7 & - & 208.9 & Cyg X-3(Binary-BH) &  \\
159 & 1MAXIS~J2044-106 & 311.05 & -10.68 & 6.0 & 6.5 & 2.0 & 3.7 & Mrk 509(AGN) & 2MAXI J2044-106 \\
160 & 1MAXIS~J2051+308 & 312.77 & 30.80 & 98.5 & 5.8 & 56.9 & 2.9 & Cygnus Loop(SNR) &  \\
161 & 1MAXIS~J2053+314 & 313.27 & 31.47 & 103.0 & 3.3 & 86.6 & - & Cyg Loop(SNR) &  \\
162 & 1MAXIS~J2103+456 & 315.95 & 45.67 & 3.0 & 6.1 & - & 6.1 & V407 Cyg(Binary-WD) &  \\
163 & 1MAXIS~J2110+466 & 317.63 & 46.67 & 5.9 & 1.0 & 4.6 & - & Cygnus Super Bubble(SNR) &  \\
164 & 1MAXIS~J2130+121 & 322.54 & 12.17 & 22.3 & 22.6 & 11.6 & 26.9 & M 15/4U 2127+119(Binary-NS) & 2MAXI J2130+122 \\
165 & 1MAXIS~J2144+382 & 326.18 & 38.29 & 149.9 & 224.5 & 289.7 & 990.8 & Cyg X-2(Binary-NS) & 2MAXI J2144+383 \\
166 & 1MAXIS~J2253+167 & 343.41 & 16.75 & 6.7 & 3.4 & 3.9 & - & IM Peg(Star) & 2MAXI J2253+166 \\
167 & 1MAXIS~J2256-030 & 344.11 & -3.07 & 2.8 & 6.4 & - & 2.9 & AO Psc(Binary-WD) & 2MAXI J2255-030 \\
168 & 1MAXIS~J2302+589 & 345.53 & 58.96 & 12.4 & 3.0 & 10.5 & - & 2E 2259.0+5836(Binary-NS/BH) &  \\
169 & 1MAXIS~J2323+588 & 350.83 & 58.81 & 59.6 & 61.1 & 74.6 & 130.4 & Cas A(SNR) &  \\
170 & 1MAXIS~J2355+286 & 358.92 & 28.67 & 6.4 & 1.2 & 2.5 & - & II Peg(Star) & 2MAXI J2354+286 \\

\end{longtable}
\end{landscape}



\end{document}